\documentclass{article}
\usepackage{graphics}
\title{Fat API bindings of C++ objects into scripting languages}
\author{Russell K. Standish\\High Performance Coders}
\newcommand{\EcoLab}{{\sffamily\slshape
    \mbox{\raisebox{.5ex}{Eco}\hspace{-.4em}{\makebox[.5em]{L}ab}}}}

\begin{document}
\maketitle
\begin{abstract}
  A {\em fat API} exposes nearly all of a C++ object's public attributes and
  methods to a consuming environment, such as a scripting language, or
  web client. This can be contrasted with a conventional, or {\em
    thin} API, where the API is defined up front, and the C++ object
  provides the implementation, most of which is private to the C++
  layer.

  Obviously, reflection is required to expose C++ objects to a
  consuming layer like this --- this paper explores using the
  Classdesc system to implement reflection of C++ objects into a
  JavaScript/TypeScript environment via a REST service, and also via a
  Node.js API module.
\end{abstract}

\section{Introduction}

Minsky\cite{Minsky} is a {\em systems dynamics}\cite{Forrester2007system} simulation package,
with an orientation towards economics, that has been under continual
development since 2011. It is implemented in C++, and historically the
user interface was implemented using the TCL/Tk toolkit\cite{Ousterhout94},
with C++ bindings provided by the EcoLab\cite{Standish01b,EcoLab} library.

From 2019-2021, the TCL/Tk layer was completely reimplemented in
Type\-script\cite{Cherny2019programming,Goldberg2022learning}, on top of
the Angular\cite{Green2013angularjs} and
Electron\cite{kredpattanakul2019transforming} toolkits, running in the Node.js\cite{ihrig2014pro}
interpreter. The advantages to doing this include accessing a much
larger ecosystem of 3rd party components, a much larger pool of
programmers (JavaScript is consistently in the top 10 of programming
languages according to the Tiobe index\cite{Tiobe}), and potentially longer term
an in-browser version of the code could be enabled via technologies
such as WebASM\cite{haas2017bringing}.

This paper reports on the subtask of exposing the Minsky's C++ core to
the TypeScript layer, allowing C++ objects to be manipulated in a
seamless manner in TypeScript code. The approach is quite general, and
could be readily adapted to other language binding APIs, or even
without an explicit binding API by means of a REST service that can be
accessed with an HTTP client implementation.

\section{REST service}

REST (REpresentational State Transfer)\cite{fielding2000architectural}
is based on web technologies. The part of a URL after the domain such
as \verb+http://www.somewhere.com/path/to/page+ is called the URL's
{\em pathinfo}. In REST terminology, it is called an {\em endpoint},
and represents a resource. What to do with the resource is given by
the HTTP verb of the request. A web browser typically performs a GET
request when you type a URL into its address bar, but there are verbs
covering all of the {\em CRUD} operations (create, read, update and
delete):

\begin{description}
\item[POST] create an object at the resource location
\item[GET] read an object at the resource location
\item[PUT] update the object
\item[DELETE] destroy the object
\end{description}

In something like an \EcoLab{} model, or the Minsky project, there is
a global static object that holds the state of the model. In the C++
code, this is accessible via a Meyer singleton pattern, ie the
\verb+minsky()+ function. So for example, a REST GET call on
\verb+/minsky/t+ returns the value of the current timestep of the
Minsky model, and performing a PUT, with floating point data in the
HTTP request body updates the timestep to the supplied value. For
convenience, the Minsky REST service ignores whether a PUT or GET is
used --- using the presence or absence of HTTP body data to determine
whether the operation is an update or a read.

One can also map method calls into the same schema. For example
\verb+/minsky/reset+ calls the reset method, which has no
arguments. The above schema for reading or updating an attribute could
be considered an example of calling an implied overloaded getter/setter
method, with overload resolution determined by the presence or
absence of data in the request body. Since we're targeting the
JavaScript ecosystem, it is natural to use JSON\cite{JSON} to encode
the parameters being passed, and the return value. Compound objects
can be serialised to/from JSON using Classdesc's existing JSON
serialiser into a JSON object (delimited by braces). Calling a method
with more than one parameter can be achieved by placing the JSON
representation of the arguments in a JSON array, which conveniently
are allowed to be of different types. So the command to export a \LaTeX{}
document describing the model's differential equation, which has
signature
\verb+void latex(const std::string& fileName, bool wrapLaTeXLines)+,
can be called through the REST service as
\verb+/minsky/latex ["foo.tex", true]+, where the first space
delineates the pathinfo and request body.

Whilst JSON is used for data encoding in this example, it is perfectly
possible to use alternate encodings. The
\verb+RESTProcess_t+\footnote{Released in Classdesc 3.43, available
  from https://classdesc.sourceforge.net, or https://github.com/highperformancecoder/classdesc.}
{\em descriptor}\footnote{In the Classdesc reflection system\cite{Madina-Standish01}, a {\em
    descriptor} is an overloaded set of function definitions that is
  mostly automatically generated by the Classdesc processor for each
  type used in the program} object has a method:
\begin{verbatim}
REST_PROCESS_BUFFER RESTProcess_t::process
  (const std::string& pathinfo, const REST_PROCESS_BUFFER& body);
\end{verbatim}
where \verb+REST_PROCESS_BUFFER+ is a macro representing the
``buffer'' concept, which defaults to \verb+json_pack_t+. A buffer implements:
\begin{itemize}
\item \verb+REST_PROCESS_BUFFER::operator>>(T&)+ for deserialisation
  to an arbitrary type
\item \verb+REST_PROCESS_BUFFER::operator<<(const T&)+ for
  serialisation of an arbitrary type
\item \verb+RESTProcessType REST_PROCESS_BUFFER::type()+ which refers
  to the type of the object serialised in the buffer
\item \verb+REST_PROCESS_BUFFER::Array REST_PROCESS_BUFFER::array() const+ returns a sequence concept object (eg std::vector or
  std::deque) if called on a \verb+REST_PROCESS_BUFFER+ that is an array, or
  usually an empty sequence if
  not. \verb+REST_PROCESS_BUFFER::Array::operator[](size_t)+ returns a
  \verb+REST_PROCESS_BUFFER+. 
\end{itemize}

The \verb+RESTProcess_t+ type is a map, where the keys are the endpoints of
the fat API, and the values are wrappers around the C++ object, or
method. These wrappers are polymorphic, with different implementations
depending on whether it is an object or a method, smart pointer or
container type. The interface is

\begin{verbatim}
  class RESTProcessBase
  {
  public:
    virtual ~RESTProcessBase() {}
    /// perform the REST operation, with \a remainder being the query string and \a arguments as body text
    virtual REST_PROCESS_BUFFER process(const string& remainder, const REST_PROCESS_BUFFER& arguments)=0;
    /// return signature(s) of the operations
    virtual REST_PROCESS_BUFFER signature() const=0;
    /// return list of subcommands to this
    virtual REST_PROCESS_BUFFER list() const=0;
    /// return type name of this
    virtual REST_PROCESS_BUFFER type() const=0;
  };
\end{verbatim}

The reason \verb+REST_PROCESS_BUFFER+ is a macro rather than a
template argument, is because \verb+RESTProcessBase+ is polymorphic,
and C++ does not allow templated virtual functions.

The methods \verb+signature+,\verb+list+ and \verb+type+ provide a
modicum of introspection to allow exploration of the fat API from the
calling side. \verb+signature+ returns an array containing the return
type and types of all arguments. 

\section{Node.js API}

Minsky's C++ layer renders directly to a native window for performance
reasons. Electron's BrowserWindow class has a native window handle
getter method that can be used to pass the native window to the C++
layer. The strategy described in the previous section of making the
C++ implementation a REST service worked well for Windows, where the
native window handles are system wide, and X-Windows system, which is
distributed by design, but unfortunately failed for the MacOSX
architecture. It turns out that Mac native window handles are actually
pointers which are, of course, only meaningful within the same process
address space.

So  the C++ layer needed to be implemented as a dynamic library, and
linked within the Node.js process using the Node.js API. Conceptually, this
is quite simple, implementing a single Node.js API endpoint (call) that
takes the pathinfo and body arguments as above. Of course, it hasn't
stayed simple --- the Node.js API allows for callbacks into the JavaScript
world from C++, which is important for some interactive functionality;
as well as also allowing offloading of C++ processing to a separate
thread, and returning the results via a JavaScript {\em promise}, which is
important for not blocking the user interface during long-running
backend operations.

\section{Attributes and Methods}

We map C++ public attributes to an implied pair of overloaded setter/getter
methods. If an argument is provided to the method, a setter is called,
and the argument assigned to the attribute. For the Minsky project,
JSON encoding of the attribute is performed, using the existing
\verb+json_pack+ and \verb+json_unpack+ descriptors. 
  
This is a very simple example of a method overload. However, C++
provides for overload resolution based on types as well as number of
arguments. JavaScript does not provide for overloaded functions at
all, but with type introspection built into the language, it is
possible to write a method that can dispatch to different
implementations based on types and number of arguments. However, with
an impoverished set of types compared with C++, this leaves us with
the problem of how to match a particular JavaScript call with a C++
method.

The approach taken in this work is to walk the C++ argument list for
each overloaded C++ method (Classdesc has been able to address
overloaded methods since version 3.37\cite{standish2019c++}), and add a
penalty for each argument that doesn't quite match. For instance if
the JavaScript environment passes a number with a non-zero fractional
part, then an integer argument C++ will receive a small penalty, but a
float or double parameter does not. If there are fewer arguments
passed than the arity of the function, or no meaningful conversion
possible, then an infinite penalty is applied. Default C++ arguments
are not supported as is, but a default argument can be reimplemented
as an overloaded method the fewer argument calls, delegating to the
method with the full number of arguments.

Finally, the method with lowest finite penalty is called, if it is
unique. Otherwise and exception is thrown back to the JavaScript
environment.

Modern C++ variadic templates are used to walk the C++ type arguments
to determine the penalty values. Then to call the C++ method, {\em
  currying} is used. The JSON arguments are converted to the relevant
C++ type, starting from the last argument, currying the bound method
to an $n-1$ argument functor, where the last argument has been fixed
by the converted JSON argument. It takes one walk through the C++
argument list to generate the curry functors, then the final zero
argument curried functor is called, which in turn calls the curried
functors up into the final bound method. The technique works well,
except that each of these curried functors need to be linked, blowing
up the build time. In \S\ref{build times}, I describe a number of
techniques to reduce the build times.

\section{TypeScript}

JavaScript, being a dynamic language, only checks numbers and types of
arguments at runtime. TypeScript\cite{Cherny2019programming, Goldberg2022learning} is an extension of
JavaScript with type annotations that are checked at compile time. For
larger more complex projects like Minsky, the TypeScript compile step
is an invaluable means of eliminating logic errors.

The JavaScript interface to C++ is of the form
\begin{verbatim}
  call("method.name", args...);
\end{verbatim}
which performs type checking at runtime. For Minsky, we created
another {\em descriptor} that outputs a series of TypeScript
definitions. This is not the only viable method. The REST API has
sufficient introspection built in, that it should be possible to build
a TypeScript script that queries the REST API, and emits the
TypeScript definitions. However doing it as a C++ process for the
Minsky project was chosen due to greater familiarity with that
environment.

For example, the \verb+Minsky+ class has a \verb+t+ double precision
attribute, a complex attribute \verb+model+ of type \verb+Group+ and
\verb+classifyOp+ method, amongst others. The custom TypeScript
descriptor outputs a definition like:
\begin{verbatim}
export class Minsky extends CppClass {
  model: Group;
  constructor(prefix: string){
    super(prefix);
    this.model=new Group(this.$prefix()+'.model');
    ...
  }
  async classifyOp(a1: string): Promise<string> 
    {return this.$callMethod('classifyOp',a1);}
  async t(...args: number[]): Promise<number> 
    {return this.$callMethod('t',...args);}
  ...
}
\end{verbatim}
The TypeScript class \verb+CppClass+ provides a number of features,
including the \verb+$prefix()+ accessor and the \verb+$callMethod()+
method that arranges for the named C++ method to be called on a
separate thread, and returns a {\em promise} that is {\em resolved} or
{\em rejected} with the return value or exception from the C++
method. Calling into C++ asynchronously in this way prevents the C++
code from blocking the GUI interface if the C++ method takes a long
time to run (as some do). There is also a \verb+$callMethodSync()+
which calls into C++ directly on the Node.js thread, which is useful when you
need to call C++ from a non-asynchronous function --- such as at
application startup. Note the use of the \verb+$+ character in the
identifier, which is a valid character in JavaScript identifiers,
but not C++, so preventing any possibility of a name clash with C++ identifiers.

To use the class definition for any object, you just have to declare:
\begin{verbatim}
  let minsky=new Minsky("minsky");
\end{verbatim}

Then you can access the time attribute via \verb+minsky.t()+ or set
the time attribute via \verb+minsky.t(10.2)+. For the complex object
\verb+model+ above, because one can call methods on it (eg
\verb+minsky.model.numItems()+), and in TypeScript identifiers cannot
be both attributes and methods at the same time, setting and getting that
object has to be done via the special \verb+$properties()+ method, ie
\verb+minsky.model.$properties()+ returns a JavaScript object
containing the public attributes of \verb+minsky.model+, and
\verb+minsky.model.$properties(object)+ sets the public attributes of
\verb+minsky.model+ using the data contained in \verb+object+.

Since \verb+minsky+ is a global object, this definition is already
provided in the backend module. But for example, the attribute
\verb+minsky.canvas.item+ is a polymorphic type with base type Item --- it can be cast to
the correct type in TypeScript via (eg)
\begin{verbatim}
  let variable=new VariableBase(minsky.canvas.item);
\end{verbatim}
then \verb+variable+ gets all of the additional attributes and methods of the
VariableBase subclass.

\section{Python}

A Python API descriptor already
exists\cite{standish2019c++}. However, it has a couple of serious
downsides. The first is that it requires the boost-python library,
which is not available currently for the MXE cross compiler\cite{MXE}, and may
never be, as it depends on the Python library being available, the
codebase of which is not friendly towards cross compilation.

The second issue is just calling the Python descriptor on the minsky
global object was not sufficient to create all the types required, and
that additional explicit descriptor calls were required to generate
all the types. This is not insurmountable --- something like this
approach was done with the TypeScript descriptor, but given the full
fat API was available through the RESTService descriptor, it was
decided to use the existing RESTService API descriptor, and write a
Python interface using the low level Python C API. That way, we should
be able to load the built Python module dynamic library into an
unmodified running Python interpreter on Windows. As well as that,
there would be no inconsistencies between the TypeScript API and the
Python API.

It was relatively straight forward, following online tutorials, to
implement a ``call'' function that takes one or two arguments, the
first being the REST function name, and the second being a JSON5
string for arguments. The second step involved creating a
REST\_PROCESS\_BUFFER object (called a PythonBuffer) that directly
marshals Python objects into their C++ counterparts without going via
JSON serialisation. Of course, for simplicity, and to avoid creating
yet another descriptor, complex objects (structs, classes etc) will
always go via JSON serialisation. Unfortunately, this exposed a
weakness in the macro approach outlined above, and the explicit
instantiation of templates, which meant that at link time there was a
definitional conflict between REST\_PROCESS\_BUFFER being a JSONBuffer
and a PythonBuffer. So for now, the PythonBuffer containing the
arguments is serialised to JSON before being passed to the
RESTProcess, and the returned JSON string used to instantiate a
PythonBuffer. Another attempt at implementing a template solution of
the RESTProcess descriptor is planned.

Finally, for return values, the PythonBuffer stores the value as an
appropriate Python object (PyObject) for the type, whether number,
string, array or so on. For objects, a custom object is returned that
has the JSON string returned by the RESTProcess stored as the
attribute \verb+_properties+ (\verb+$+ is not a valid character in
Python identifiers), and also new callable attributes for each method,
allowing usage like
\begin{verbatim}
  r=container._elem(2).method()
\end{verbatim}
within Python code.

\section{Build time optimisation}

\label{build times}

As previously alluded, extensive use of variadic templates for
processing overloaded functions caused a dramatic impact on compile
times for the Minsky project, which went from circa 2 minutes for the
TCL/Tk version (which doesn't support overloaded methods) to around 20
minutes for the JavaScript build. Profiling the build times indicated
a massive increase in the time taken to link the ``executable'' --- in
this case a dynamic library with a \verb+.node+ extension that Node.js
loads as an ``add on''.

One of the identified reasons for the slowdown in linking speeds is
the large number of generated template helper functions to handle
introspection of functional objects. The number grows as the square of
the number of arguments of the method, and linking objects is
$O(n^2)$, so the link time grows as the 4th power of the number of
method arguments. As noted later, the link times for standard Linux
linkers is not actually too bad --- in the few years since this work
was started, Linux linkers have improved remarkably.

In some way, the link strategy is quite stupid, as these helper
functions only need to be used on one place in one object file, and so
resolved at compile time. This suggested a strategy of privately
declaring the variadic templates and explicitly instantiating them
within just a single object file where they were used ---
unfortunately, the compiler still emitted symbols for each and every
helper template, even if they're not linked to from other object
files, and this technique didn't help.

So the next thing was to remove the RESTProcess '.rcd' definition
files from the include headers, and include them in just one
compilation unit, and explicitly instantiate the template within that
compilation unit. This improved the build time quite significantly.

The next strategy tried, is to do things the old-fashioned
way. Instead of recursively defined variadic templates, explicit
templates created by means of a shell script that creates explicit
support functions for 0, 1, 2 etc arity functions up to some
predefined maximum value (6 was found to be the maximum arity function
present, with the renderWindow method being one of the biggest).

The final strategy was to reduce the maximum arity of the exposed
methods. The simplest way to do this, given that one could pass a
Javascript object which is packed and then unpacked into the C++
object via JSON, is to rollup several of the arguments into a compound
object. In this way, the maximum arity was reduced to 4.

Finally, it turned out that the clang ecosystem had a much more
performant compiler and linker for these purposes than the GCC
ecosystem, and that template unrolling gave negligible benefit in the
clang case.

\begin{table}
  \begin{center}
    \begin{tabular}{r|cc}
      Strategy& GCC & Clang \\\hline
      None & 1048 & 377 \\
      Explicit instantiation & 445 & 287 \\
      Unrolled templates & 427 & 291\\
      Arity reduction & 409 & 284 \\
    \end{tabular}
    \caption{Build times for the different build time optimisations
      for the two different compiler toolchains.}
    \label{table:build times}
  \end{center}
\end{table}

\begin{figure}
  \resizebox{\textwidth}{!}{\includegraphics{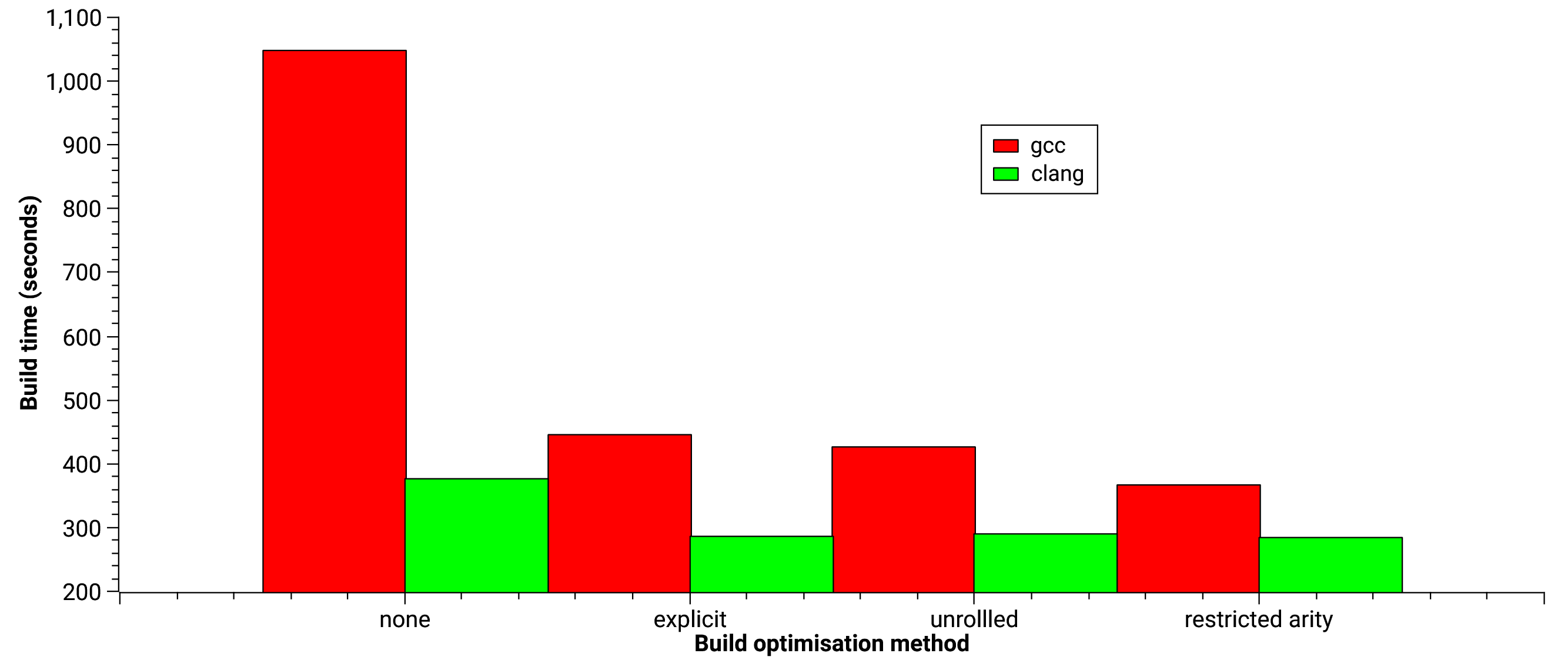}}
  \caption{Build times for the different build time optimisations
    for the two different compiler toolchains.}
  \label{fig:build times}
\end{figure}

Table \ref{table:build times} shows the build times for the various build time
optimisations described in the text above, displayed graphically in
figure \ref{fig:build times}. The optimisations were
    applied consecutively from top to bottom, so that the unrolled
    template method was applied to explicitly instantiated code, and
    so on.

The final test was to try the extremely performant {\em
  mold}\cite{mold} linker. As per Mold's README, adding the flag
\verb+-fuse_ld=mold+ is sufficient to delegate the link step to mold. Link
times were measured by building the target (minskyRESTService.node),
removing just the target, leaving all the object files present, and
timing how long it takes to build the target again. 

\begin{table}
  \begin{center}
    \begin{tabular}{r|r|l}
      Linker & Version & Time (seconds) \\\hline
      GNU ld & 2.41 & 4 \\
      LLVM ld (lld) & 15.07 & 3.9 \\
      Mold & 2.3 & 0.7 \\
      MXE ld.bfd & 2.37 & 791 \\
    \end{tabular}
    \caption{Link times for various linkers tested}
    \label{link times}
  \end{center}
\end{table}

As can be seen from table \ref{link times}, for Linux builds, the
linking time is inconsequential, well within noise, so even though
Mold is blazingly fast, there is no particular advantage for this
project. What isn't inconsequential is the link time for generating
Windows versions of the Node.js addon, which takes over 13 minutes. Just
quite why the linker is so slow for Windows is unclear, however a neat
trick discovered whilst doing this benchmarking is to symbolically
link the LLVM linker \verb+ld.lld+ to the MXE linker
\verb+x86_64-w64-mingw32.shared-ld+. It works just as well, and only takes
around 4 seconds.

\section{Methods}

Build times were recorded using the inbuilt ``time'' command, running
on a quad-core Intel(R) Core(TM) i5-1135G7, at 3.8GHz, with a Samsung
970 EVO 500GB NVMe M.2 SSD. The operating system was OpenSUSE Leap
15.5, and the compilers used: GCC 13.2.1 and Clang 15.0.7.

The codebase used was Minsky 3.3.2,\footnote{Available from
  https://minsky.sourceforge.net, or
  https://github.com/highperformancecoder/minsky} except for the
``none'' strategy above. In explicitly instantiating the templates
that define the descriptor, it is not feasible to put the code change
behind a feature flag. Going back to the earlier version of the code
will not be comparing apples with apples, as about a year's worth of
development has occurred since that change. So the particular
optimisations were backed out from the 3.3.2 codebase: the explicit
instantiations removed (they were implemented in a macro, so this was
easy), then the inlined descriptor definitions included back in the
header files. The code changes were committed to the branch
compile-optimisations-undone\footnotemark.

Particular optimisation feature flags can be turned on via Makefile
flags, as shown in table \ref{make-commands}. The command was run
after an initial \verb+make -j9+ to ensure all prerequisites were
built, to avoid including the prerequisites build time. One can
measure the overhead time required for make to start up via \verb+make -n+, which
proved to be about 1.3 seconds, so well within experimental noise.

\begin{table}
  \begin{center}
    \begin{tabular}{l|r}
      Toolchain,Strategy & Command \\\hline
      GCC,none\footnotemark[\thefootnote] &\verb+rm *.i; time make -j9 GCC=1 CLASSDESC_ARITIES=+\\
      Clang,none\footnotemark[\thefootnote]&\verb+rm *.i; time make -j9 GCC= CLASSDESC_ARITIES=+\\
      GCC,explicit&\verb+rm *.i; time make -j9 GCC=1 CLASSDESC_ARITIES=+\\
      Clang,explicit&\verb+rm *.i; time make -j9 GCC= CLASSDESC_ARITIES=+\\
      GCC,unrolled&\verb+rm *.i; time make -j9 GCC=1 CLASSDESC_ARITIES=0xffff+\\
      Clang,unrolled&\verb+rm *.i; time make -j9 GCC= CLASSDESC_ARITIES=0xffff+\\
      GCC,arity reduction&\verb+rm *.i; time make -j9 GCC=1 CLASSDESC_ARITIES=0xf+\\
      Clang,arity reduction&\verb+rm *.i; time make -j9 GCC= CLASSDESC_ARITIES=0xf+\\
      Link time & \verb+rm gui-js/node-addons/minskyRESTService.node; \+\\
      GCC link time&\verb+time make -j9 GCC=1+\\
      Clang link time&\verb+time make -j9 GCC=+\\
      Mold link time&\verb+time make -j9 OPT=-fuse_ld=mold+\\
    \end{tabular}
  \end{center}
  \caption{Commands for timing different optimisation strategies.}
  \label{make-commands}
\end{table}

\footnotetext[\thefootnote]{compile-optimisations-undone branch, available from
  https://github.com/highperformancecoder/minsky}

\section{Conclusion}

The RESTService API descriptor provides a scripting language
independent fat API interface to C++ code. Method arguments and return
values can be marshaled using a custom native type ``buffer'' object,
or using JSON5 encoding with the preexisting Classdesc json
descriptor. In practice, JSON5 encoding tends to be sufficiently
performant. Both a Javascript and Python bindings were generated
automatically for the Minsky systems dynamics simulator, and
furthermore, TypeScript binding were generated automatically though a
custom descriptor, leading to easier to read scripting code, and
relatively more type-safe use in Minsky's front end code.

Using the RESTService descriptor comes at additional build cost,
compared with the original TCL bindings used for the \EcoLab{}
package, which is ameliorated via a number of C++ coding techniques,
the use of the Clang toolchain over the GCC one, and the use of modern
Linux linkers.

\bibliographystyle{plain}
\bibliography{rus}

\end{document}